\documentclass[twocolumn,showpacs,preprintnumbers,floatfix]{revtex4}
\usepackage[dvips]{graphicx}
\usepackage{amsmath,amssymb}
\begin{document}
\preprint{123XXXX}
\title{\textbf{On the Corporate Votes and their relation with Daisy Models}}
\author{H. Hern\'andez-Salda\~na}
\email{hhs@correo.azc.uam.mx }
\affiliation{Departamento de Ciencias B\'asicas,\\
Universidad Aut\'{o}noma  Metropolitana Azcapotzalco,\\
Av. San Pablo 180, M\'{e}xico 02200 D.F., Mexico.}
\date{October, 2008}
\begin{abstract}

The distribution of votes of one of the corporate parties in Mexico during elections of 
2000, 2003 and 2006 is analyzed.
After proper normalization and unfolding, the agreement of the votes distributions with 
those  of  daisy models of several ranks is good. These models are generated 
by retaining each $r+1$ level in a sequence which follows a Poisson distribution.
Beyond the fact that 
rank $2$ daisy model resembles the distribution of the quasi-optimal 
distances for the Traveling Salesman Problem, no clear explanation exists for this behavior,
but the agreement is not fortuitous and the possibility of a universal phenomena for corporate vote 
is discussed.

\end{abstract}

\pacs{89.75.-k, 87.23.Ge}
\maketitle

\section{Introduction}

Several recent efforts have been done in order to understand the subyacent dynamics in electoral 
systems and opinion formation \cite{TailsTies,Castellano}, from the contrarian effect\cite{Galam} to the, so called, small 
world behavior\cite{SmallWorld,Herrmann}. In the Brazilian elections, for instance, power law was found for the proportional 
vote\cite{CostaFilho1999,CostaFilho2006}.
However, to establish  a proper description of an electoral process is a hard issue since many 
factors and interactions appear and several aspects of them must be studied. 
Statistical characterization of actual processes is an important issue as well, mainly with 
the increasing possibility of obtaining the  vote data.
In the present work we  incorporate the analysis on the corporate vote with a study on the statistical properties of
the federal Mexican elections of 2000(E-2000), 2003(E-2003), and 2006(E-2006). Since their distributions are 
smooth, the existence of an analytical distribution that describes them and a model which explains them, are  
very tempting issues. We were successful in the first topic but the answers to the second 
remain open. We find a  
remarkable well fit of the properly unfolded distribution of votes with a family of distributions obtained in 
the context of spectral statistics of complex quantum systems, the called daisy models~\cite{hhs}.
The process presented here is different from those  that appears in \cite{Castellano}, for instance,
since the vote decision is taken due to pertain to a corporate.

In the referred electoral processes  two new features appeared: i) the party 
who ruled for around $70$ years\cite{PRI} became opposition and ii) the vote data  are 
public and in an electronic format\cite{IFE, given}. The last  fact allows an extense statistical analysis, 
meanwhile, the former one gave the opportunity  to analyze the vote distribution of the 
hard core or corporate voters. We shall denote this party as P2 according to the place in which 
it appears in the data basis\cite{PRI}. 

Here we present the vote distribution  of this party during the elections 2000, 2003, and 2006 for 
the president position and places in both chambers. This distribution consist of the histogram of the 
number of votes per cabin obtained for each party, i.e., in how many cabins exists $1$ vote, $2$ votes and so on.
The crude data histograms are presented in the next section,  and the unfolding procedure together with the 
theoretical distribution appear in section \ref{Unfolding} followed by some remarks and conclusions.

\begin{figure}[!tb]
\begin{center}
\includegraphics[width=1.0\columnwidth,angle=0,scale=1.0]{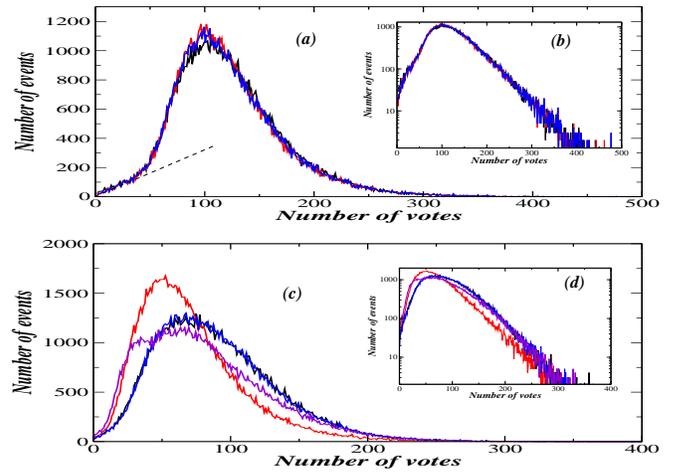}
\caption{\footnotesize{(Color online) Histogram of the number of votes obtained in each cabin for the 
P2 party\cite{PRI} during
the electoral processes of (a) 2000 and (c) 2006 for deputies (black), presidential (red) and senators (blue), and
the deputies from the 2003 election ((c) violet). In the inserts (b) and (c) appear the corresponding log-linear graphs
which show the exponential decay in all the cases. Note that the distributions in (a) start with a linear behavior
and the dashed line is for guide the eye. No average of any sort was considered.
} }
\label{Fig:1}
\end{center}
\end{figure}

\section{The data}
In a recent analysis of the Mexican election of 2006\cite{Baezetal} the distributions of votes for 
all the participants were done with data obtained from the preliminary
results program (PREP).
For the two main parties the results are unclear and a mix of processes is expected; meanwhile 
for the rest of the forces clear distributions appear: power laws for small parties, annulled votes and non-registered
candidates, and a smooth distribution for the third electoral force. 
The corporate party P2 present a histogram with a clear maximum and a tail with exponential decay.
In Fig. ~\ref{Fig:1}(a) the histograms  of the official final results for P2 are presented. The 
data were obtained from the electoral authorities web page \cite{IFE} and on request \cite{given}.
By construction, each cabin admits only $750$ votes from the registered list of voters and they are distributed
over all the country  being a sample grid on the population aged over $18$ years old\cite{Vcard}. 
As an exception, there exist special cabins for voters in transit but 
the number of them is small and do not affect the statistical results presented here. With those remarks, 
it is clear that the histograms are statistics. The analysis of the link with the geo-economic regions
is beyond of the present work but it is of interest.

In Fig.~\ref{Fig:1}(a) and ~\ref{Fig:1}(c) the histograms for presidential (red), deputies (black) and senators (blue)
for E-2000 (upper panel) and E-2006 (lower panel) are presented and the intermediate elections for the 
low chamber in E-2003 (Fig.~\ref{Fig:1}(c) in violet). We present the crude data with no average of any sort. 
Other parameters of the distributions are presented in Table \ref{Table:1} being the total number of  cabins
considered in each process the following:  $113423$ for E-2000, $121367$ for E-2003 and $125962$ for 
E-2006. The 
existence of a smooth distribution that fits the data is a very tempting issue and is the matter of the rest
of the present work.

\begin{table}
\begin{tabular}{|l|r|r|r|} 
\hline
 Year &Total of votes& Avg.& S.D. \\
\hline
Dep. 2000 & 13734103 &  121.130  &  51.331 \\ \hline
Pres. 2000 & 13575704 & 119.695 & 50.414 \\ \hline
Sen. 2000 & 13618056  &  120.095 & 50.908 \\ \hline
Dep. 2003 & 9878787 & 81.461 &  46.700 \\ 
\hline
Dep. 2006  & 11339480  & 90.023 & 44.166 \\ \hline
Pres. 2006 & 8960369   & 71.136 & 38.599 \\ \hline
Sen. 2006  & 11292853  & 89.653 & 43.920 \\ \hline
\end{tabular}
\caption{\footnotesize{Total number of votes obtained by  P2 at several elections. The average (Avg.) and 
standard deviation (S.D.) correspond to the arithmetical ones.}}
\label{Table:1} 
\end{table}

\section{Unfolding and Fitting}\label{Unfolding}

\begin{figure}[htb]
\includegraphics[width=1.0\columnwidth,angle=0,scale=0.85]{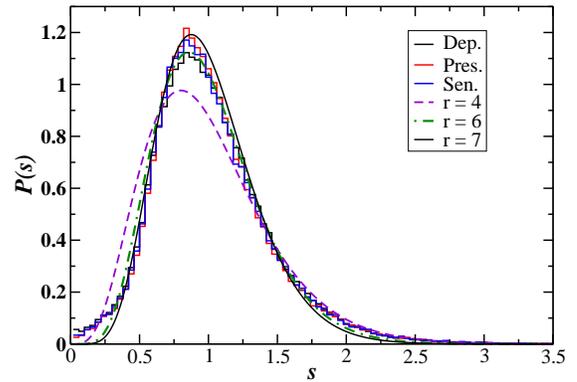}
\includegraphics[width=1.0\columnwidth,angle=0,scale=0.85]{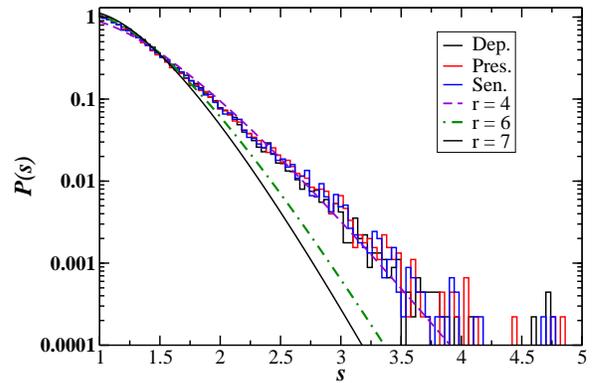}
\caption{\footnotesize{(Color online) Unfolded distribution of votes for E-2000 in histograms. The
continuous lines correspond to daisy model of Eq. \ref{rdaisy} with $n=1$ and $r=4$, $6$, and $7$ as
indicated in the inset box.
} }
\label{Fig:2}
\end{figure}

A direct comparison with any probabilistic distribution  requires of proper normalization and 
unfolding of the signal, i.e.  separate the secular part from the fluctuating one.  
To consider this procedure is important since in many cases considering the relative variable 
$x/< x >$ is not enough because the 
average $<x>$ could not be constant through the whole set of data. 
In general, 
the unfolding procedure is a very delicate task \cite{WeidenmullerGuhr}.
In the present case no 
{\it a priori} density can be defined, since it is not clear if the alphabetical order in which the data basis is 
ordered corresponds to the dynamics of the system. To test the fitting, two sorts were considered, the original
order and a randomly sorted sequence of the votes. In both cases the histograms are similar but the last one gives
much more stable results after the unfolding procedure, as we shall discuss below.

In order to fit the experimental data with a probability distribution we treat
the number of votes $n_i$ as if they were differences of energies in a quantum spectrum. As in the case of 
energy levels, we consider the spectrum,$\{x_i\}$, formed by levels 
\begin{equation}
x_{i+1} = x_i + n_{i+1},
\label{intvote}
\end{equation} 
where $n_i$ is the 
number of votes in the cabin $i$ and we define $x_1 = n_1$. It is costumary to consider the  integrated spectral 
function or integrated density $ {\cal{N}}(x) = \sum_i \Theta ( x - x_i) $, which counts the number of levels 
$x_i$ with value equal or less that $x$. $\Theta(\cdot)$ stands for the Heaveside unitary step function.  
The integrated density is decomposed into a secular $\tilde{\cal{N}}$ and a fluctuating part ${\cal{N}}_{fluc}$.
The former part is given by the integral  of the correlation function of one point (See Ref.~\cite{WeidenmullerGuhr}
for explanation). The sequence $x_i$ is mapped onto the numbers $y_i$, with $y_i = \tilde{\cal{N}} (x_i)$. 
The new variable is the unfolded one which has a constant density and the statistical analysis is
performed on it. In the case of quantum systems $\tilde{\cal{N}}$ is estimated applying semiclassical rules. 
The first term of its expansion is called, in the  literature, the Thomas-Fermi estimate or, the Weyl term 
in the case of billiards. In practical situations this function is evaluated via polynomial fitting. 

\begin{figure}[t]
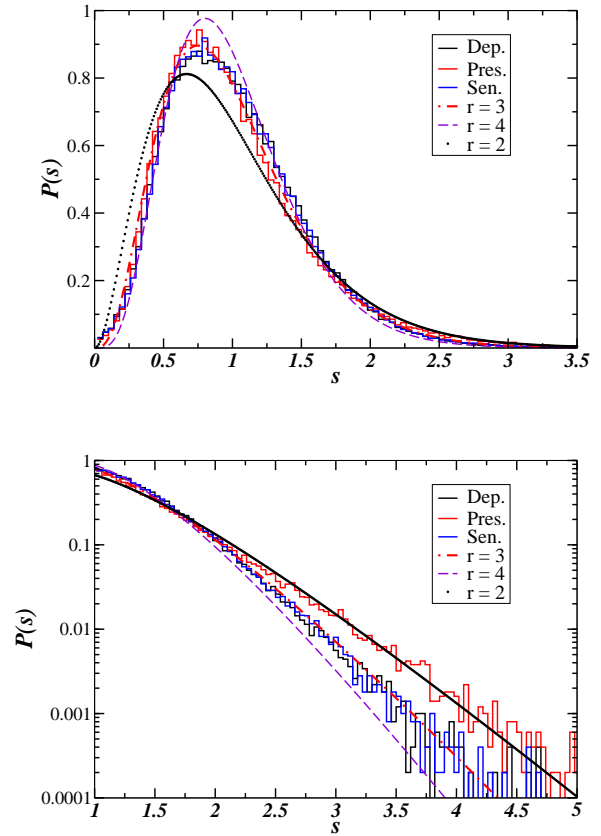

\begin{center}
\includegraphics[width=1.0\columnwidth,angle=0,scale=0.85]{pri2006.eps}
\includegraphics[width=1.0\columnwidth,angle=0,scale=0.85]{pri2006log.eps}
\caption{\footnotesize{(Color online) Same as before but for E-2006. 
} }
\label{Fig:3}
\end{center}
\end{figure}

The last procedure described was done in the present case  since it is the standard in many fields of
physics and it is of general applicability. The specific statistic that we analyze  
corresponds to the nearest neighbor spacing, $s$, defined as $s_{i+1} = y_{i+1} - y_{i}$ and 
is the unfolded version of the number of votes $n_i$ defined in Eq. (\ref{intvote}).
The results for the unfolded votes $s_i$ are shown in Figs. ~\ref{Fig:2} and ~\ref{Fig:3}  
for the deputies, president
and senators in  2000 and 2006 elections. For sake of clarity we drop the E-2003 analysis.

The theoretical comparison is made with the daisy model of rank $r$~\cite{hhs}. This model departs
from retaining each $r+1$ level from a set of levels with a Poisson distribution. The resulting sequence has the 
$r+1$-nearest neighbor distribution of Poisson's, but it must be renormalized in order to obtain 
the nearest neighbor distribution of the daisy model of rank $r$. The resulting $n$-th neighbor spacing distribution is 
\begin{equation}
P_r(n,s)=\frac {(r+1)^{(r+1)n}}{\Gamma([r+1]n)}s^{(r+1)n-1}\exp[-(r+1)s].
\label{rdaisy}
\end{equation}
where $r$ corresponds 
to the kind or rank of the family. 
The rank $r=1$ corresponds to the Semi-Poisson distribution which is related to 
the energy distribution of pseudointegrable systems\cite{prosen} and others systems at 
criticality\cite{Bogomolny,Garcia} like as the disordered conductor at the Anderson transition
\cite{Shklovskii}. A strong relation exists between daisy models and the nearest-neighbor interaction
one-dimensional Coulomb gas\cite{Bogomolny, Gerland}, where the dependence in the inverse temperature $\beta$ from the
later model has the same role as the rank $r$ in the former.

We do no have an {\it a priori} density in order to compare the vote records with the distribution of 
Eq. (\ref{rdaisy}). Then, the present study can be done only at
nearest neighbors, $n=1$, even when an exploration to larger range correlations 
is extremely interesting and will be matter of future works. 

The model described by Eq. (\ref{rdaisy}) with $n=1$ fits the unfolded  P2 distribution of vote in 
two regions with two different daisy ranks, $r$,  the
central part is usually fitted by a higher rank and decays with a lower one. In the case of E-2000 
(Fig. \ref{Fig:2}) the 
fit is between  $r=6$ and $r=7$, but a remarkable deviation is that the experimental distributions 
start with a linear grow as indicated below and with the dashed line in Fig.~\ref{Fig:1}(a). This 
characteristic remains after the unfolding procedure and marks a clear deviation from the 
$\sim s^r$ behavior, however, if we do not consider this data, the area preservation of the 
distribution makes that the function with $r=7$ fits better. The decay is  well fitted by a $r=4$ daisy model in all 
the cases, as can be seen  in the lower panel of Fig. \ref{Fig:2}. It is important to note that a 
fit to Weibull/Brody distribution is not good, since the decay is clearly exponential and that 
only happens in the Poissonian case of such distribution. Clearly this is not the case.

In E-2006 a differential vote occurs since the presidential candidate obtains 
around $20\%$ less votes than P2 obtains for the chambers, as can see in Table \ref{Table:1}. Such an event 
does not happens in the E-2000 process. In this case, the vote distributions for the chambers fit with 
a $r=3$  for the whole range, the body and the tail (Fig. \ref{Fig:3}). For the presidential case exists 
a general fit with the $r=3$ for the body, but decays with a rank $2$. Note that the agreement 
for the presidential case (red histogram) is not as good as the other cases. The main difference
in E-2000 and E-2006 is that P2 arrived to the last process with a deep internal division as 
was widely reported in national newspapers. 

An interesting remark, since the votes distribution has fit different rank of daisy models is that 
the parameter $r+1$ plays the role of an inverse temperature when the these models are contrasted 
with the statistical distributions of a $1$-dimensional Coulomb gas with logaritmic 
interactions~\cite{hhs,Bogomolny} 

\section{Conclusions}

In this work we presented an analysis of the vote distribution for one of the 
corporate parties (named P2 here) in Mexico which has a wide  influence in all the country since 
it was in the federal power for around $70$ years. By construction the Mexican electoral system admits
a straight statistical analysis since the number of cabins are defined and distributed in order that each
cabin admits only $750$ voters.
The crude P2 vote distributions look smooth (Fig. \ref{Fig:1}) and, after a proper 
unfolding and normalization, corresponds to a probability distribution. Comparison of 
the data with nearest neighbor daisy models of rank $r$ ($n=1$) gives a good agreement 
in all the cases. In the  election in 2000 the agreement could be better, but the 
data distribution depart from  a linear grow as indicated in Fig.~\ref{Fig:1}(a) with a dashed line tor guide the 
eye. The distributions tails follow a different daisy model rank.

The dynamical meaning of these results is unclear but the agreement with a daisy model suggest the 
existence of universal processes therein and not just a fortuitous agreement. 
In Ref. \cite{hhs} the daisy model of rank $2$ fit the 
distribution of distances for the quasi-optimal path in the Traveling Salesman Problem (TSP). The problem
consists in finding the shortest path between $N$ cities visiting each  city just one time. It is clear 
that the problem presented here is of the same type. 
P2 have voters in each region of the country and 
their conform a truly wide web. How this happens is matter of future analysis, as well the existence of 
similar behavior in other corporate parties around the world. 

\section{Acknowledgment}
This work was partially supported by DGAPA-UNAM project IN-104400 and PROMEP 2115/35621.


\begin{thebibliography}{oo}
\bibitem{SmallWorld} A large amount of existing references can be obtained in Science Citation Index.
\bibitem{given} According with the mexican transparency law, the electoral data are public and can be obtained on request.
\bibitem{Vcard} To be registered in the voters list is almost mandatory in Mexico since 
is required as the official ID card for all practical purposes.
\end{thebibliography}
\end{document}